# Ultra-broadband near-field Josephson microwave microscopy

Ping Zhang[1,+], Yang-Yang Lyu[1,+,*], Jingjing Lv[1], Zihan Wei[1,2], Shixian Chen[1], Chenguang Wang[1,2], Hongmei Du[1], Dingding Li[1], Zixi Wang[1], Shoucheng Hou[1], Runfeng Su[1], Hancong Sun[2], Yuan Du[1], Li Du[1], Liming Gao[3], Yong-Lei Wang[1,2,*], Huabing Wang[1,2,*], Peiheng Wu[1,2]

[1]*School of Electronic Science and Engineering, Nanjing University, Nanjing 210023, China*

[2]*Purple Mountain Laboratories, Nanjing 211111, China*

[3]*Institute of Electronic Materials and Technology, School of Materials Science and Engineering, Shanghai Jiao Tong University, Shanghai, 200240, China*

+ Authors contribute equally

**\*Email:** yylyu@nju.edu.cn, yongleiwang@nju.edu.cn, hbwang@nju.edu.cn.

## Abstract

**Advanced microwave technologies constitute the foundation of a wide range of modern sciences, including quantum computing[1-3], microwave photonics[4,5], spintronics[6,7], etc. To facilitate the design of chip-based microwave devices[8,9], there is an increasing demand for state-of-the-art microscopic techniques capable of characterizing the near-field microwave distribution and performance. In this work, we integrate Josephson junctions onto a nano-sized quartz tip, forming a highly sensitive microwave mixer on-tip. This allows us to conduct spectroscopic imaging of near-field microwave distributions with high spatial resolution. Leveraging its microwave-sensitive characteristics[10,11], our Josephson microscope achieves a broad detecting bandwidth of up to 200 GHz with remarkable frequency and intensity sensitivities. Our work emphasizes the benefits of utilizing the Josephson microscope as a real-time, non-destructive technique to advance integrated microwave electronics.**

# Main text

Microwave technology, based on precise generation, manipulation, and measurement of microwaves, plays a pivotal role across various fields, including circuit quantum electrodynamics[1-3], microwave photonics[4,5], spintronics[6,7], etc. These devices are primarily reliant on precise microwave control, of which advancements often depend on the exploration of nanoscale microwave materials and underlying mechanisms. Meanwhile, these integrated devices often demand efficient microwave transmission with supporting planar micro-circuits[8,9], where the integration poses challenges to electromagnetic incompatibility and signal crosstalk. For purposes of achieving optimal performance, there is a growing need to detect near-field microwave distribution from microwave devices, aiming to analyze mechanisms of microwave interaction and facilitate efficient signal coupling.

Presently, several existing microwave imaging techniques have been developed, including coplanar waveguide probes[12,13], scanning near-field microwave microscopy[14-16] (SNMM), nitrogen-vacancy (NV) center microscopy[17-19], atomic vapor cells[20,21], etc. Conventional coplanar waveguide probes and SNMM, for instance, rely on the metal-coated tip to facilitate microwave coupling and transmission along RF cables. The microwave loss during this process will become more severe with the decrease of tip size, as well as the increase of microwave frequency, so it is difficult to balance the sensitivity and spatial resolution when detecting the weak microwave emission from the device surface. High-sensitivity microscopy, taking NV-center microscopy as the most representative technique, typically requires external magnetic fields to broaden their frequency band, especially when there is a need to measure microwave signals up to several tens of gigahertz[19]. The biasing magnetic fields can substantially degrade magnetic-sensitive devices and their associated circuits. Consequently, there's an urgent requirement for the development of a high-sensitive and field-free microwave microscopy technique.

The Josephson junction refers to a structure where two superconductors are separated by a thin insulator or a normal conductor, which is an important element in superconducting electronics. It constitutes a weakly coupled junction where Cooper pairs can tunnel through the barrier via the tunneling effect, traversing from one side to the other[22]. The profoundly nonlinear characteristics of Josephson current can be used as mixers[23], with their sensitivity approaching the quantum noise limit[24]. Furthermore, when exceeding the characteristic frequency of Josephson junctions, it can be used as bolometers[25] and exhibits exceptional sensitivity to microwave[26], even capable of detecting single microwave photons[10,11]. However, conventional on-chip Josephson junctions cannot be brought in close proximity to devices at the limit, thus making it challenging to achieve near-field imaging with high spatial resolution. Previous works fabricated superconducting quantum interference devices (SQUID) onto nanoscale tips to achieve high-spatial-resolution detection of static magnetic field[27,28] and thermal phenomena[29,30], providing a new idea to explore the microscopic physical phenomena through the utilization of Josephson junctions.

Here, we develop an innovative Josephson microscope by integrating Josephson junctions directly onto a nano-probe, thereby enabling efficient low-temperature near-field microwave characterization (Fig. 1a). The Josephson microscope demonstrates a broadband coherent detecting frequency up to 200 GHz, as well as a power sensitivity of -75 dBm and a spatial resolution of submicron. Notably, passive detection ensures non-destructive imaging and *in-situ* characterization without external biasing fields, thereby keeping the normal operational state of devices under test (DUT), especially for highly sensitive devices.

Serving as the key component of microscope, the Josephson probe is fabricated by depositing niobium (Nb) films onto a nano-sized quartz tube using DC magnetron sputtering (Extended Data Fig. 1). The deep grooves on the tube separate Nb films, thus forming weak links at the apex (left panel of Fig. 1b). Traditional probe fabrication utilizes directional evaporations and uses electron beam evaporation to deposit Nb film[31], which has such a high melting temperature that common instruments without sufficient cooling can't handle. This innovative approach of utilizing magnetron sputtering overcomes the challenges related to isotropic deposition, achieving Nb films with high qualities. The characterization system of the Josephson microscope consists of DC and AC modules (right panel of Fig. 1b). Detailed descriptions can be found in the Methods section.

## Microwave sensing

The transport performance of the Josephson probe is first demonstrated to evaluate the capability of microwave sensing. Here, we take probe #1 as a demonstration. The probe has a superconducting critical temperature ($T_{c0}$) of about 5 K and exhibits non-hysteretic current-voltage (*I~V*) characteristics (Extended Data Figs. 2a-b), which is a typical feature of superconducting Dayem-bridges. All measurements were conducted at a bath temperature of 3.3 K. Under microwave irradiations, the Josephson probe demonstrates resonance phenomena. This resonance is characterized by discrete, step-like constant currents in its *I~V* characteristics[32], denoted as Shapiro steps (Fig. 2a). The voltage positions of Shapiro steps follow the AC Josephson voltage-frequency relationship of 2e/h = 483.6 MHz/μV (where h is the Planck constant and e is the elementary charge), while the width of Shapiro steps is relevant to microwave intensity. Shapiro steps indicate the existence of Josephson junctions at the probes' apex (Extended Data Fig. 3), confirming the ability of microwave coherent detection at the frequencies of the signal applied. The highest microwave frequency that the Josephson probe can currently detect is above 200 GHz. Taking probe #2 as another example, it also shows non-hysteretic transport characteristics (Extended Data Figs. 2c-d). Under the irradiation at each frequency, the curves of differential resistance exhibit dips at corresponding voltages (Fig. 2b), representing the Shapiro steps. The frequency range of detectable microwaves is determined by the product of the probe's superconducting critical currents ($I_c$) and normal resistance ($R_n$), which will be discussed in the frequency-tracking section.

To accurately describe the intensity resolution, we used a commercial microwave source to measure the *I~V* characteristics of probe #1. Figure 2c shows a clear visualization of the voltage response and differential resistance as a function of probe voltages and microwave powers. The observed current steps on the *I~V* characteristics correspond to the minima of the d*V*/d*I* values, following exactly the positions at integers times of h$f_{mw}$/2e, where $f_{mw}$ is 10 GHz in this case. As the microwave signal is emitted in space, there exists an enormous loss due to the spatial coupling between the RF cable and the probe. For estimation of the actual microwave intensity received by the probe, we fit *I~V* characteristics by the classical RCSJ model (Extended Data Fig. 4). The original *I~V* characteristic is fitted to obtain the Stewart-McCumber parameter of 0.1, confirming a non-hysteretic feature. After fitting the microwave response of the probe's *I~V* characteristic, the attenuation coefficient of the whole system can be calculated to be 53.6 dB. To evaluate the intensity resolution of the Josephson probe, we obtain the probe's voltage changing with the microwave power, represented as d$V$/d$P$, with the variation of bias currents (inset of Fig. 2c). The highest microwave response is observed near the probe's critical current ($I_c$ ~ 30 μA), with a maximum value up to $4.81 \times 10^4$ V/W. Then the noise voltage spectral density is tracked with a bias current around the probe's $I_c$ (Fig. 2d). The voltage noise displays a transition from low-frequency 1/$f$ noise to high-frequency white noise, with a voltage noise level of approximately 1.5 μV Hz$^{-1/2}$ in the kHz range. Taking into account the voltage noise and the probe's responsivity, the sensitivity of the probe is around -75.1 dBm (31.2 pW Hz$^{-1/2}$ @ 3.3 K). Consequently, the Josephson probe demonstrates exceptional adaptability to microwave chips with low power consumption.

## Frequency tracking

The highly nonlinear nature of Josephson junctions makes them exceptional mixers[33]. Unlike conventional Josephson junction mixers receiving signals from far field, the Josephson probe integrates a nano-sized Josephson mixer onto a tip, enabling near-field microwave detection. Specifically, this technique allows the ability of frequency tracking towards weak microwave signals based on fundamental or harmonic frequency mixing (Fig. 3a), effectively eliminating the effects of transmission losses and noise interference. When the Josephson probe receives a radio frequency ($f_{RF}$) input from the DUT, it generates a frequency mixing product with the local oscillation ($f_{LO}$) input provided by a microwave source, typically operating within a few GHz range. Subsequently, the resultant intermediate frequency ($f_{IF}$) output is amplified and collected by a spectrum analyzer. This method provides a means of calculating high-frequency $f_{RF}$ (~ GHz/THz) signals by detecting low-frequency $f_{IF}$ (~ MHz) signals (Fig. 3b). When the Josephson probe detects both $f_{LO}$ and $f_{RF}$ signals, its intrinsic nonlinearity generates $f_{IF}$ signals given by |$mf_{RF} \pm nf_{LO}$|, where $m$ and $n$ are positive integers. Here, we focus on the low-frequency output with $m$ = 1, resulting in $f_{IF}$ = |$nf_{LO} - f_{RF}$|, which is in the hundreds of MHz range and easier for amplification and detection. By capturing the corresponding $f_{IF}$, we can deduce the DUT's frequency via $f_{RF} = nf_{LO} \pm f_{IF}$, where + or – can be easily decided by the change of $f_{IF}$ with the that of $f_{LO}$.

Figure 3c presents the probe's capability of frequency tracking by Josephson mixing. To demonstrate the process, we generate a replica of an 'unknown' $f_{RF}$ signal using another microwave source. A microwave signal of 40 GHz is involved for probe #1 (Fig. 3c). For every 0.1 MHz change in the $f_{LO}$ signal, the $f_{IF}$ output shifts by approximately 1.6 MHz, indicating a 16th harmonic mixing. By iteratively adjusting the frequency of $f_{LO}$ and collating measurement results, we determined an average value of $f_{RF}$ to be precisely 40 GHz ± 2.2 kHz (Fig. 3d). Fundamental frequency mixing is also conducted with $f_{LO}$ close to $f_{RF}$, and the results present the same frequency value of 40 GHz (Extended Data Fig. 5). The counts of values are divided into only two columns, showing the limitation of our spectrum analyzer and the possibility for further optimization. For probe #2, a microwave signal is generated by a Gunn oscillator and emitted onto the probe. The frequency of the microwave is around 112 GHz, yet the accurate value remains to be measured. The $f_{LO}$ signals change with an interval of 1 MHz, while the corresponding $f_{IF}$ signals vary 8 MHz in each step (Fig. 3e). By fitting the equation of $f_{RF} = 8 \times f_{LO} - f_{IF}$, the unknown frequency is calculated to be around 111.45 GHz (Fig. 3f). Since the amplitude of $f_{LO}$ remains unchanged, the amplitude and frequency of $f_{IF}$ output only depend on those of $f_{RF}$. Therefore, we can achieve a frequency-selective intensity detection based on frequency mixing, e.g., resolving the intensity distribution of certain frequencies from resonators with various frequency modes.

The upper limit of the detectable frequency range for the probe, denoted as the characteristic frequency ($f_c$), is theoretically determined by the product of $I_c$ and $R_n$ in Josephson junctions. Combining the AC Josephson voltage-frequency parameter with the measured $I_cR_n$ value of the probes, $f_c$ can be derived by $f_c = I_cR_n \times (2e/h)$. We obtain $I \sim V$ characteristics from different batches of probes and calculate the $f_c$ values (Extended Data Fig. 6). The $f_c$ of probes is determined by the apex size of the quartz tube and the thickness of the Nb film, which can be well controlled from several GHz to 200 GHz. For tracking frequencies much lower than the probe's $f_c$, the thermal noise may destroy the synchronization between the Josephson AC supercurrents and external microwave[34,35], so that the probe's capability of coherent detection is lost. Thus, probes with different $f_c$s are recommended to be selectively employed for microwave characterization scenarios involving varying frequency ranges.

## Spatial imaging

Coplanar waveguide (CPW) constitutes an indispensable component of the microwave resonator and finds widespread application in superconducting quantum circuits[36,37]. To demonstrate the microwave imaging capabilities of the Josephson microscope, we designed a typical CPW structure composed of Nb film. The CPW is configured such that the microwave signal is fed in from the left side, while the other side remains open to create an impedance mismatch (Extended Data Fig. 7). As a result, standing wave along the central conductor always occurs with its periodic length depending on injecting microwave frequencies. We imaged the spatial distribution of these standing waves within the CPW at a height of 10 μm above the sample surface. The Josephson probe is biased around its $I_c$, and voltages were tracked and spatially plotted into a colormap (Fig. 4a). The result reveals a clear periodic distribution along the central conductor, with at least two cycles inside. The tip-sample distance is also crucial when detecting near-field microwave signals. Although the imaging height is much smaller than the microwave wavelength (tens of millimeters), the signals of microwave distribution become blurred as the tip-sample distance gradually increases from 1 μm to 100 μm (Extended Data Fig. 8), indicating the necessity of near-field imaging for accurate microwave characterization.

Conventional CPW structures typically operate at the millimeter scale, lacking nanoscale features. To showcase the spatial resolution of the Josephson microscope, we selected another structure 'capacitor', which has found its wide applications in microwave quantum technologies[38,39]. We designed a nano-sized interdigital capacitor (Fig. 4b), with a width of 500 nm for the central electrodes and 2.5 μm gaps in between. An optical photograph of the interdigital capacitor is presented in the background of Fig. 4b. Microwaves are fed from the top electrode, while the bottom electrode is open. The colored curves in Fig. 4b represent the results of line scanning along the *x*-axis under different intensities of microwave injection. As the microwave intensity increases, voltage signals detected from the probe progressively rise, allowing for clear observation of the microwave signals emitted from the electromagnetic coupling between interdigital electrodes. The peaks and dips of measured curves are consistent with both the shape and quantity of interdigital electrodes, clearly displaying sub-micron spatial features. Considering the frequency of the feed-in microwave and a resolution better than 1 μm, the Josephson microscope has reached a detecting scale of $5 \times 10^{-6}~\lambda$ so far. By further incorporating tip-sample-distance control methods commonly employed in traditional scanning probe microscopy, such as the microscopes based on tuning forks[26] and optical interferometers[40], a nanoscale spatial resolution could be feasibly attained.

## Conclusion

In summary, our study introduces a low-temperature passive near-field microwave microscopy by placing nano-sized Josephson mixers onto the probe. We demonstrate its remarkable capabilities in terms of broad bandwidth up to 200 GHz and multiple operating modes, paving the way for near-field characterization of weak signals from chip-based microwave devices. The experiments at various temperatures reveal that lower temperatures can enhance the performance of the Josephson probe (Extended Data Fig. 9), making it compatible with the environment at ultra-low temperatures that are required for quantum circuits. The Josephson probe presents an effective wideband, high-resolution mixing detection methodology, enabling passive near-field identification of subtle high-frequency signals.

Applications of the Josephson microscope into even higher frequency ranges are possible by optimizing the probe with different structures and/or materials, such as fabricating superconductor-insulator-superconductor junctions[41,42]. On the other hand, the Josephson probe can be used as a nanoscale bolometer, enabling non-coherent detection in the terahertz domain[43]. Our microscope represents an *in-situ* and non-destructive characterization tool compatible with quantum technologies operating at ultra-low temperatures. It can also directly advance the on-chip circuit design and facilitate the discovery of novel phenomena in terahertz physics[44,45,46], spintronics[47], and metamaterials[48].

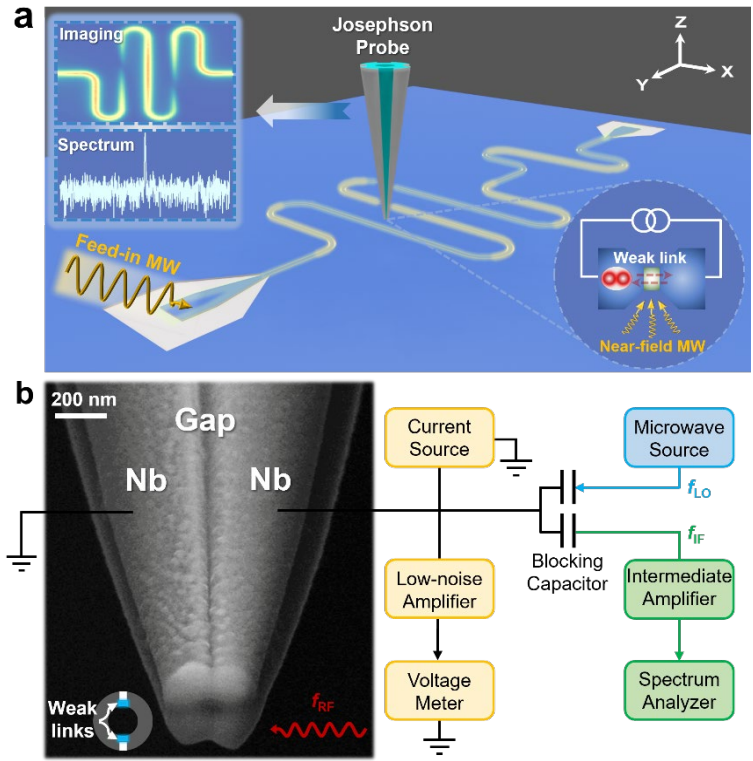

**Figure 1 | Demonstration of the Josephson microscope. a**, Microscope assembly. When the probe approaches the DUT, weak-link Josephson junctions on the apex receive subtle near-field microwaves emitted from the DUT surface, causing oscillations of Cooper pairs. The probe is biased by a current source to perform coherent detection and characterizations of microwave distribution and spectrum tracking. **b**, Measurement system. The probe's SEM image (left) shows the formation of weak-link Josephson junctions on top separated by grooves. External measuring circuits are divided into DC and AC modules separated by blocking capacitors, as illustrated on the right. The DC module employs customized low-noise source meters for probe biasing and readout. AC signals ($f_{IF}$ and $f_{LO}$) are transmitted through DC-blocking capacitors to conduct frequency mixing, when the Josephson probe receives microwave ($f_{RF}$) from the DUT.

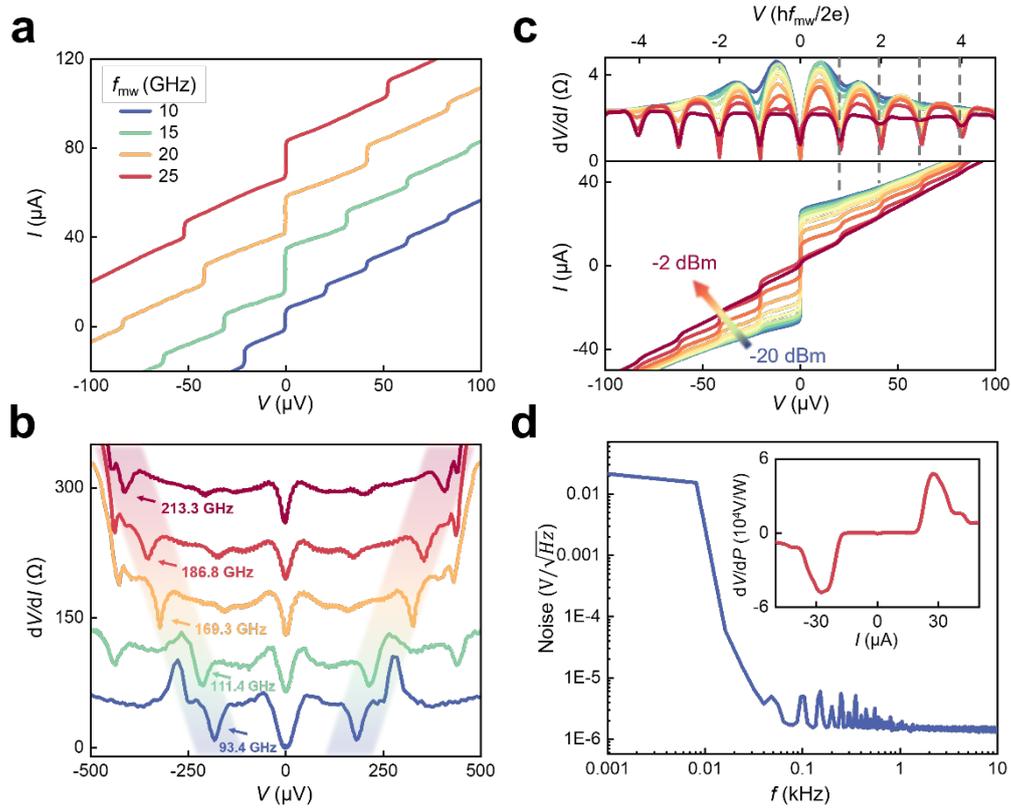

**Figure 2 | Microwave response of Josephson probes. a**, Microwave response for probe #1. Shapiro steps at different positions were presented in I~V characteristics when the probe is exposed to microwaves with frequencies of 10/15/20/25 GHz. The curves were shifted along the *y*-axis (25 μA intervals) for clarity. **b**, Microwave response for probe #2. Shapiro steps were observed as dips in the voltage dependence of differential resistances, when being exposed to microwaves above 100 GHz. The features are marked out by arrows with corresponding microwave frequencies. The curves were shifted along the *y*-axis (65 Ω intervals) for clarity. **c**, Microwave intensity dependence of I~V characteristics for probe #1. The lower graph shows the Shapiro steps when the probe is irradiated to 10 GHz microwaves with the intensity from -20 dBm to -2 dBm. The upper graph shows the voltage dependence of differential resistance corresponding to the I~V characteristics below. **d**, Noise voltage spectral density for probe #1. The probe is biased around the critical current $I_c$ and the white noise level reaches a minimum value of 1.5 μV Hz$^{-1/2}$. The probe's microwave responsivity shown in the inset is obtained by subtracting two I~V characteristics shown in Extended Data Fig. 4 and considering the attenuation coefficient. All measurements mentioned above were conducted at a bath temperature of 3.3 K.

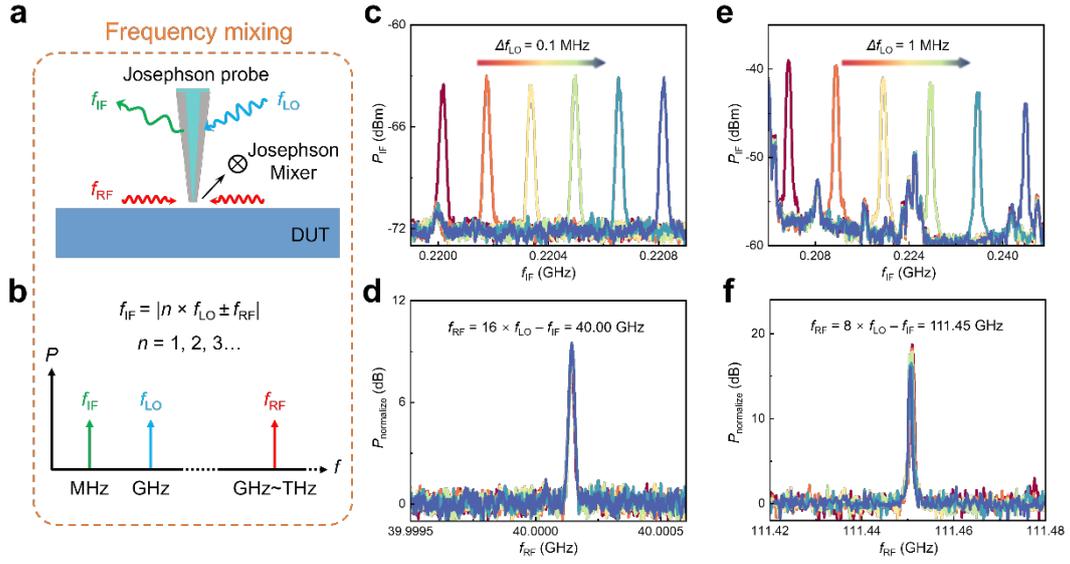

**Figure 3 | Frequency tracking by Josephson probe. a**, Principle of the frequency mixing. The Josephson probe acts as a mixer and receives $f_{RF}$ signals from the DUT surface. By interacting with a biased local oscillation at $f_{LO}$, the probe transforms them into $f_{IF}$ outputs. **b**, Frequency bands of involving signals. The frequency bands of $f_{IF}$ and $f_{LO}$ are located in MHz and GHz, respectively, while the $f_{RF}$ signal could reach up to GHz or even THz. Three signals satisfy the relationship of $f_{IF} = |n \times f_{LO} \pm f_{RF}|$. **c**, Harmonic frequency mixing for probe #1. The provided $f_{LO}$ is changing from 2.51372 GHz to 2.51386 GHz with an interval of 0.1 MHz (from left to right). Peaks of the spectra indicate that the corresponding $f_{IF}$ outputs shift by an interval of 1.6 MHz, indicating $\Delta f_{IF} = 16 \ast \Delta f_{LO}$. **d**, Restored spectra of $f_{RF}$. By regarding $f_{LO}$ as a single frequency output and fitting the equation of $f_{RF} = 16 \times f_{LO} - f_{IF}$, the spectra of $f_{RF}$ could be exactly reconstructed from (**c**). **e**, Harmonic frequency mixing for probe #2. Peaks of the spectra indicate the shift of $f_{IF}$ output is 8 MHz in each step, while $f_{LO}$ is changing from 13.9568 GHz to 13.9618 GHz with an interval of 1 MHz (from left to right). **f**, Restored spectra of $f_{RF}$. By fitting the equation of $f_{RF} = 8 \times f_{LO} - f_{IF}$, the unknown $f_{RF}$ is estimated to be around 111.45 GHz.

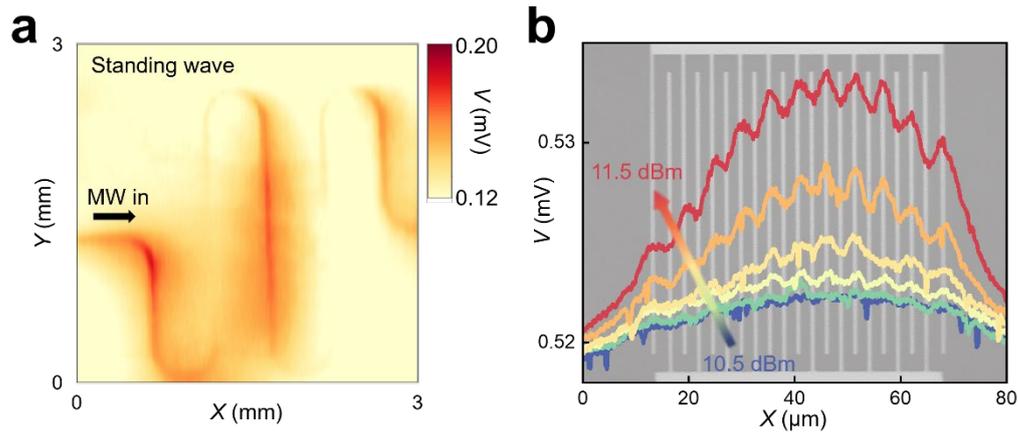

**Figure 4 | Spatial imaging by the Josephson microscope. a**, Intensity distribution of CPW. The signals are obtained by tracking the probe's voltage with a bias current around $I_c$ while scanning the CPW with an injecting microwave of 12.28 GHz. The complete structural design of CPW is shown in Extended Data Fig. 7. **b**, Line scan signals across the nano-capacitor. The background shows an optical photograph of the nano-capacitor, with the *y*-axis distorted for better visualization. The colored lines represent the probe voltages corresponding to different microwave input intensities with a frequency of 1.4 GHz.

# Method

**Probe fabrication.** The fabrication process involves a quartz tube with a diameter of 1.2 mm and four deep grooves spaced at 90-degree intervals on the sidewall. Laser heating during the pulling process caused the tube to fracture into two (Sutter Instrument P-2000), resulting in a nano-tip with broken nodes reaching an approximate diameter of 100 nm. The deep grooves around the tip remained intact. Subsequently, Nb films were deposited using DC magnetron sputtering (AdNaNoTek Corp.) in a specific sequence: the first deposition was performed with the probe positioned perpendicular to the direction of plasma, the second deposition involved rotating the probe 180 degrees axially, and the third deposition was carried out parallel to the plasma. Josephson junctions were formed as weak links at the apex of the probe.

**Microscope setup.** All transport characteristics were measured in a helium-free cryostat with low vibration (Montana S100). The probe is capsuled in a homemade shielding holder with low-pass filters and mounted on the cold head of the cryostat. The scanning capability of the microscope is provided by three-axis positioners (Attocube ANPx51 and ANPz51). DC signals from the probe were obtained using a traditional four-terminal configuration. The bias current was supplied by a custom-made current source powered by two 12 V batteries to minimize interference. Output signals were amplified by a low-noise amplifier before being collected by a data acquisition card (NI PCI-6221). The required microwave input, serving as the local oscillation signal, was coupled to the probe by a commercial microwave source (Agilent N5183A). The high-frequency microwave signals were provided by a Signal Generator Extension module (Virginia Diodes, Inc. 80-250 GHz) or a Gunn oscillator (Epsilon Lambda Electronic Corp. ELMI94/U). The intermediate signal resulting from frequency mixing was passed through a custom-made intermediate amplifier and then sent to a spectrum analyzer (Agilent N9010A).

**Testing devices.** The coplanar waveguide and interdigital capacitor devices were fabricated using a 50-nm Nb film obtained by DC magnetron sputtering. The fabrication process involved photolithography and lift-off procedures to define the desired patterns. The central electrodes of the coplanar waveguide had a width of 30 μm, and the equivalent impedance was designed to be 50 Ω. The nano interdigital capacitor had a width of 0.5 μm for the central electrodes, with 2.5 μm gaps in between.

## Acknowledgments

The authors gratefully acknowledge financial support by the National Key R&D Program of China (2021YFA0718802 and 2018YFA0209002), the National Natural Science Foundation of China (61727805, 62288101, 62101243, 62274086 and 62201396), Jiangsu Key Laboratory of Advanced Techniques for Manipulating Electromagnetic Waves, China Postdoctoral Science Foundation and Jiangsu Outstanding Postdoctoral Program.

## Author contributions

Y.W., H.W. and P.W. conceived the idea and designed the experiments. P.Z., S.C. and S.H. fabricated the probes. J.L. and Z.W. conducted SEM characterization. P.Z., C.W., Z.W., H.D. and D.L. performed the measurements. R.S., H.S., Y.D., L.D. and L.G. discussed the results and fitting. Y.L. and Y.W. drafted the manuscript, with contributions from all authors. P.Z. and Y.L. contributed equally to this work.

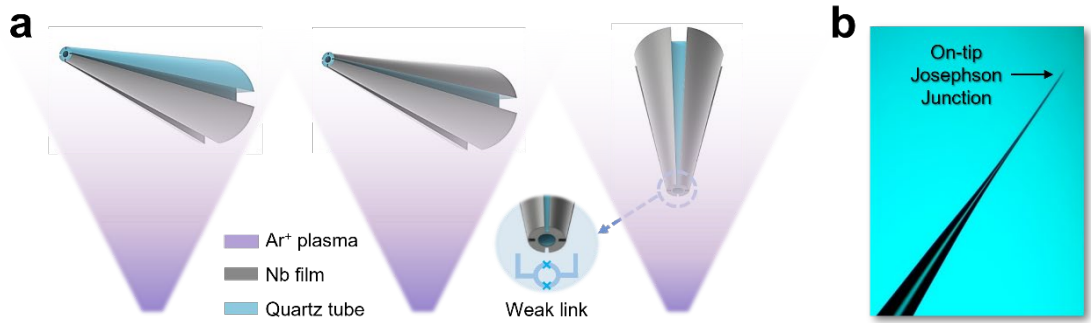

**Extended Data Figure 1 | Demonstration of Josephson probes. a**, Fabrication procedure. Niobium film (grey) was deposited onto the quartz tube (blue) by DC magnetron sputtering following three rounds: firstly, the probe is placed perpendicular to the plasma direction; secondly, it's axially rotated 180 degrees; thirdly, the probe is rotated parallel to the plasma direction, leading to the formation of weak-link Josephson junctions at the apex. **b**, Optical photo. Josephson junctions are located on the apex of the probe.

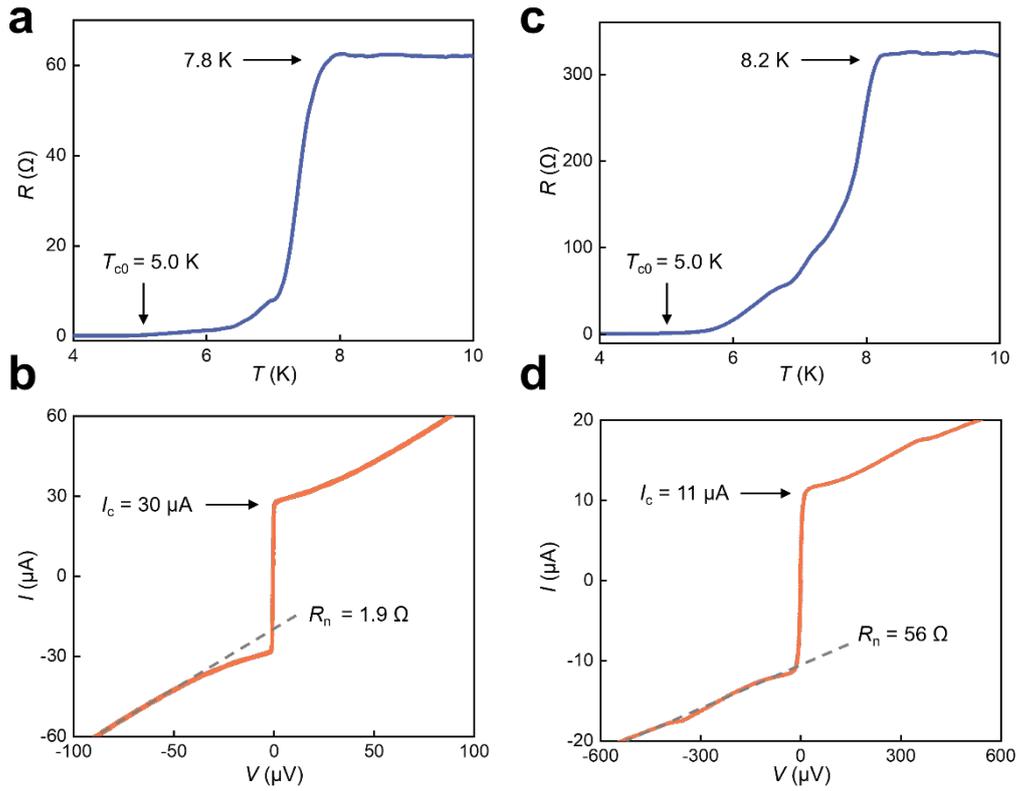

**Extended Data Figure 2 | Transport properties of Josephson probes. a**, Temperature dependence of resistance for probe #1. The resistance starts to drop around 7.8 K and reaches zero at 5.0 K. **b**, $I\sim V$ characteristics for probe #1. The current was sweeping in a full loop and the curve shows no hysteresis. The probe has a critical current ($I_c$) of 30 μA and a normal resistance ($R_n$) of 1.9 Ω. As shown in the graph, $I_c$ is generally defined by a threshold of 5 μV, above the voltage noise level. $R_n$ is defined by the slope of the linear region in the $I\sim V$ characteristic, where the Josephson junction loses its superconducting state. **c**, Temperature dependence of resistance for probe #2. The resistance starts to drop around 8.2 K and reaches zero around 5.0 K. A residual resistance of 8.9 Ω is subtracted. **d**, $I\sim V$ characteristics for probe #2. The probe has a critical current of 11 μA and a normal resistance of 56 Ω.

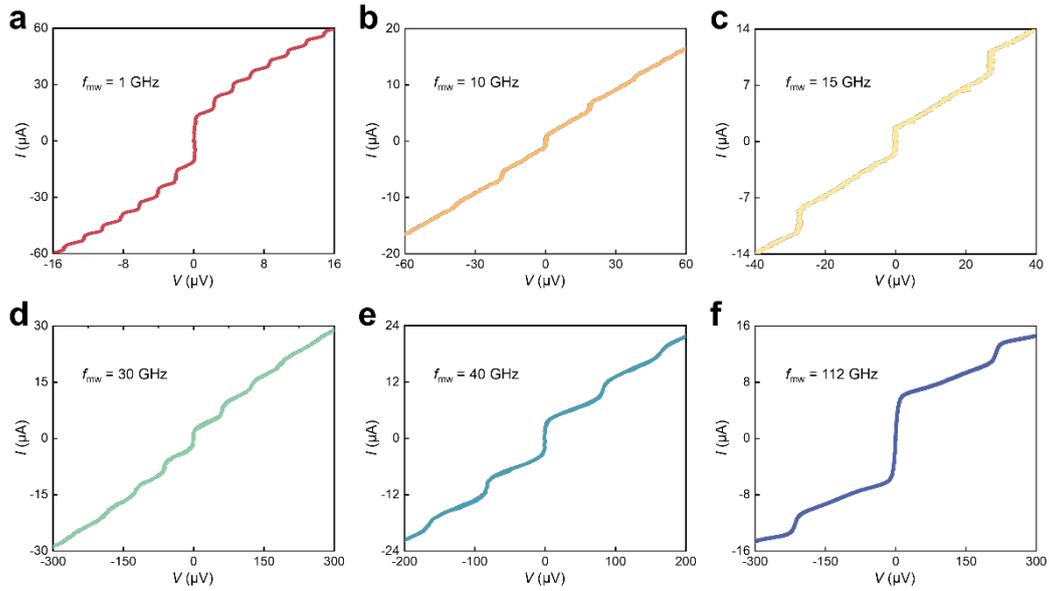

**Extended Data Figure 3 | Microwave responses of different batches of probes.** Microwave signals of 1~40 GHz are provided by a commercial microwave source, while the 112 GHz microwave signal is provided by a Gunn oscillator. The presence of Shapiro steps confirms the existence of Josephson junctions on those tips, showing a repeatable procedure to fabricate Josephson probes.

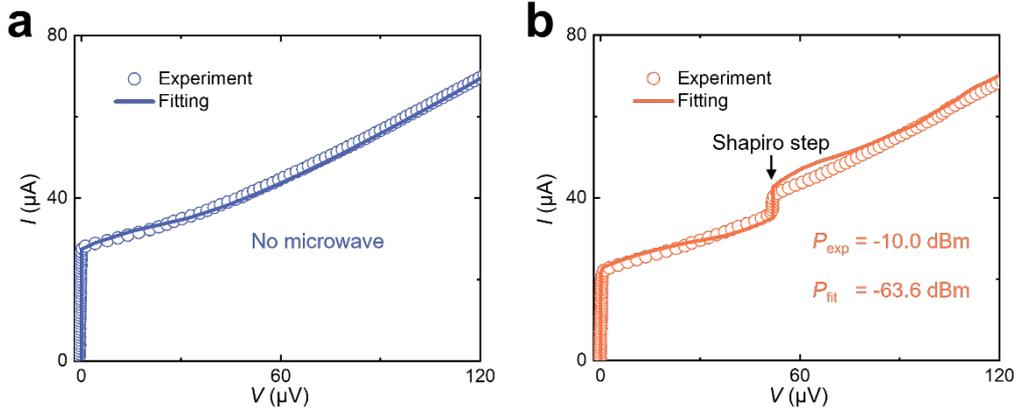

**Extended Data Figure 4 | Comparison between experiments and fitting for probe #1. a,** Fitting of original $I$~$V$ characteristic. Dots represent experimental the $I$~$V$ characteristic under -10 dBm shown in Fig. 2b, while solid lines are simulating results fitted by the RCSJ model. In a normalized form, the RCSJ model can be written as $i_b=\beta_c d^2\varphi/d\tau^2+d\varphi/d\tau+\sin\varphi$. The bias current ($i_b$) consists of $i_{dc}+i_{ac}\sin(\omega\tau)+i_n\sin(\omega_n\tau)$, corresponding to d.c. bias currents, a.c. currents and currents produced by external noise, respectively. In the formula, $\omega$ is the Josephson plasma period, $\omega_n$ is the external noise period, $\varphi$ is the Josephson phase difference and $\beta_c$ is the Stewart-McCumber parameter. **b,** Fitting of $I$~$V$ characteristic at -10 dBm. As $I_{ac}$ reaches 21.3 µA, the fitting curve shows the same features as that in the experimental data under a microwave intensity of -10 dBm. The mismatching between experiment and simulation may result from thermal noises, ambient noise, or excess currents. Combining the probe's $R_n$ of 1.92 Ω, the microwave intensity theoretically acting on the probe is around -63.6 dBm.

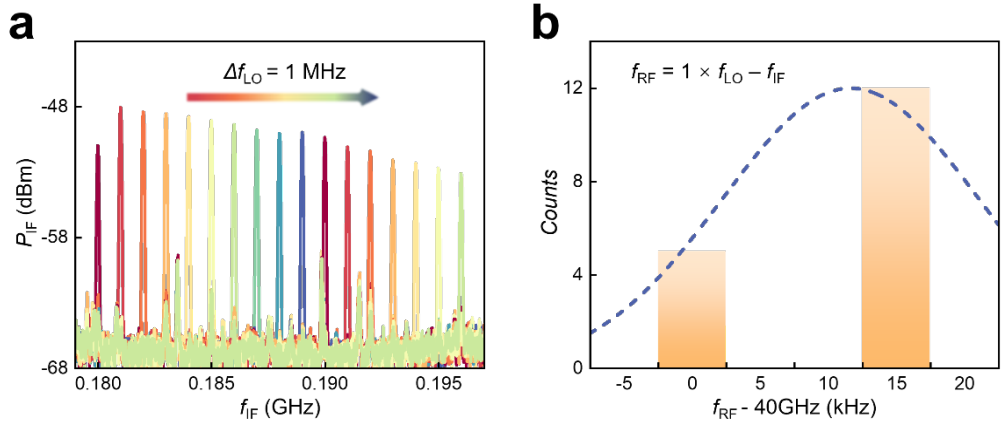

**Extended Data Figure 5 | a**, Fundamental frequency mixing for probe #1. Peaks of the spectra indicate the variations of $f_{IF}$ output, while $f_{LO}$ is changing from 40.180 GHz to 40.196 GHz, with an interval of 1 MHz and 17 times in total (from left to right). **b**, Calculated $f_{RF}$ values. By fitting the equation of $f_{RF}$ = 1×$f_{LO}$ - $f_{IF}$ from the spectrum in (**a**), the frequency of $f_{RF}$ could be reconstructed to be exactly 40 GHz. The $f_{RF}$ in the x-axis is subtracted by a value of 40 GHz.

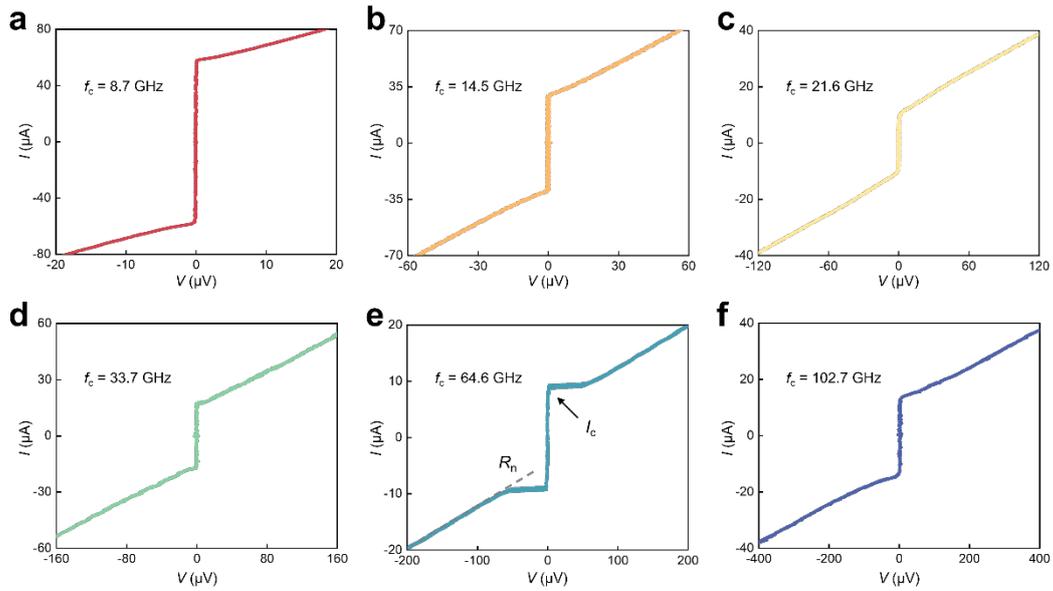

**Extended Data Figure 6 | Transport properties of different probes.** The bias current is sweeping in a full loop and $I$~$V$ characteristics normally show no hysteresis. The critical currents ($I_c$) of probes typically range from 1 μA to 100 μA. The upper limit of frequency detection ($f_c$) shown on the left-top corner can be calculated by $f_c = I_c R_n \times (2e/h)$, indicating the frequency range where these probes can conduct coherent detection. As shown in (e), $I_c$ is generally defined by a threshold of 5 μV, above the noise level. The normal resistance ($R_n$) is defined by the slope of the linear region in the $I$~$V$ characteristic, where the Josephson junction loses its superconducting state. Due to certain reasons like excess currents, the defined $R_n$ is slightly higher than its intrinsic $R_n$, but it is sufficient to estimate the value of $f_c$.

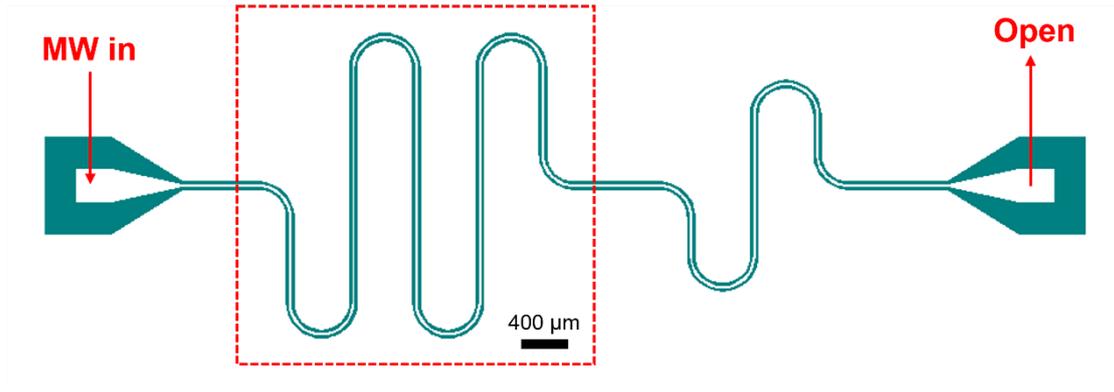

**Extended Data Figure 7 | Design of the coplanar waveguide (CPW). a**, Design diagram of CPW. The central conductor of the CPW has a width of 30 μm and an equivalent impedance of 50 ohms. The microwave signal is fed in from the left side, while the right side remains open-circuited. The red frame marks out the scanning region, which reaches the maximum area (3×3 mm$^2$) offered by the scanning platforms.

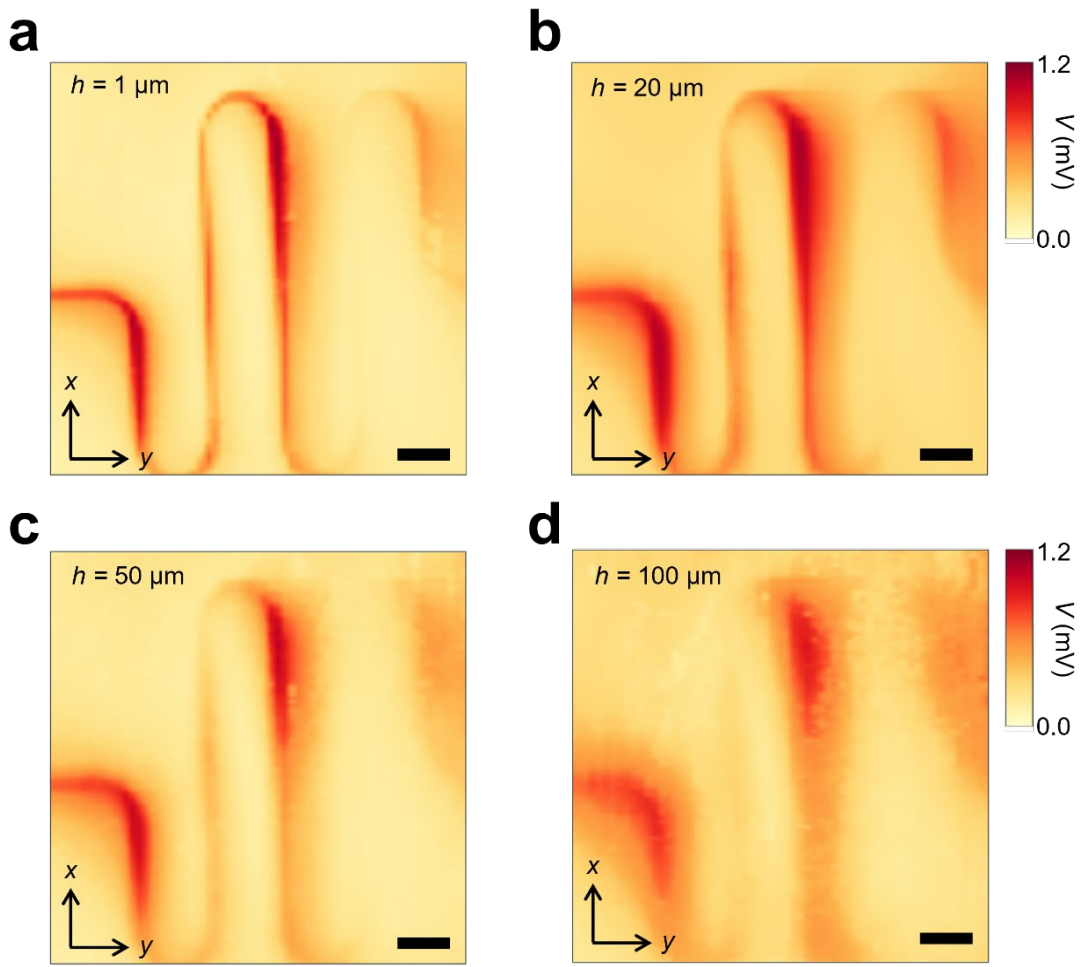

**Extended Data Figure 8 | Intensity distribution of CPW on different heights.** The signals are obtained by tracking the probe's voltages while scanning the CPW. The injecting microwave is at 2.90 GHz. The tip-sample distances ($h$) vary from 1 μm to 100 μm in (**a**)-(**d**). All color bars are shown in the same range for comparison. Scale bar, 400 μm.

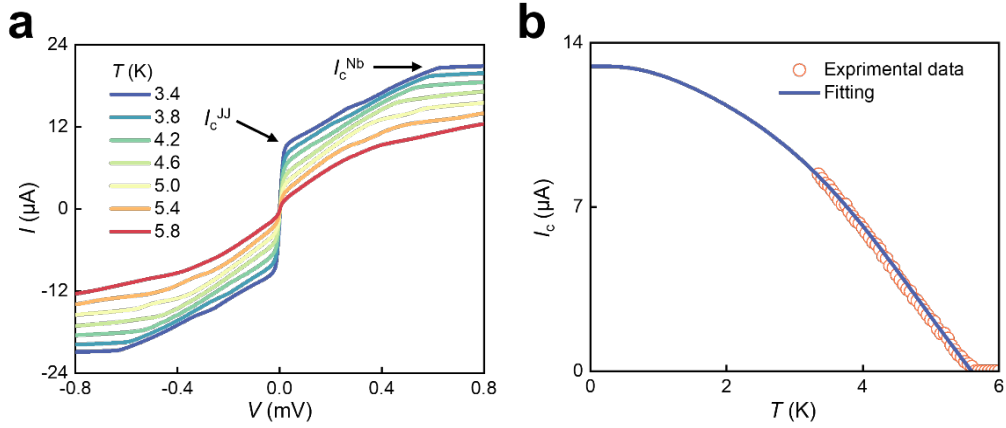

**Extended Data Figure 9 | Temperature dependence of the probe. a**, $I\sim V$ characteristics at different temperatures. The critical currents of the Josephson junction ($I_c^{JJ}$) and Nb film ($I_c^{Nb}$) are marked out in the graph. **b**, Temperature dependence of the probe's $I_c^{JJ}$. Red dots represent the probe's $I_c$s extracted from (**a**), by defining the probe's $I_c$ at a differential resistance of 5 Ω. The blue dashed line indicates the fitting curve based on the Ambegaokar-Baratoff formula $J_c(T) = \frac{\pi\Delta(T)}{2eR_n}\tanh\frac{\Delta(T)}{2k_BT}$, where $\Delta(T)$ is the superconducting energy gap, $R_n$ is the normal resistance of the junction, $e$ is the electron charge and $k_B$ is the Boltzmann constant. The temperature dependence experiment reveals that the $I_c$ of the junction increases with decreasing temperature, reaching the maximum at millikelvin. Since $I_cR_n$ defines the probe's detecting responsivity and upper limit of detecting frequency, the increased $I_c$ effectively raises the probe's performance at ultra-low temperatures.